\documentstyle[aps,prb,twocolumn,amsfonts]{revtex}
\input{psfig}
\newcommand{\Tr}{\mathop{\mbox{Tr}}\nolimits}
\newcommand{\diag}{\mathop{\mbox{diag}}\nolimits}
\def\N{{\bf N}}
\def\n{{\bf n}}
\def\bsigma{{\bbox{\sigma}}}
\def\wbar{{\overline{w}}}

\def\U{{\rm U}}
\def\SL{{\rm SL}}
\begin{document}
\twocolumn[\hsize\textwidth\columnwidth\hsize\csname@twocolumnfalse\endcsname
\title{Counting statistics for arbitrary cycles in quantum pumps}
\author{Yuriy Makhlin$^{1,2}$ and Alexander Mirlin$^{3,4}$}
\address{$^1$Institut f\"ur Theoretische Festk\"orperphysik,
Universit\"at Karlsruhe, 76128 Karlsruhe, Germany}
\address{$^2$Landau Institute for Theoretical Physics, 
Kosygin st. 2, 117940 Moscow, Russia}
\address{$^3$Institut f\"ur Nanotechnologie, Forschungszentrum
Karlsruhe, 76021 Karlsruhe, Germany}
\address{$^4$Institut f\"ur Theorie der Kondensierten Materie,
Universit\"at Karlsruhe, 76128 Karlsruhe, Germany}
\date{\today}
\maketitle
\begin{abstract}
Statistics of charge transport in an adiabatic pump are determined by
the dynamics of the scattering matrix $S(t)$.  We show that, up
to an integer offset, the statistics depend only on the corresponding path
$\N(t)=S^\dagger\sigma_3 S$ in the coset space (the sphere for a single 
channel). For a general loop $S(t)$ we solve for the noise-minimizing pumping 
strategy. The average current is given by the area enclosed by $\N(t)$ in the 
coset space; its minimal noise by the area of a minimal surface (soap film) 
spanned by 
$\N(t)$ in the space of all matrices.  We formulate conditions for
quantization of the pumped charge.
\end{abstract}
\vspace{0mm}]
Adiabatic charge pumping has attracted considerable attention recently, largely
motivated by the experiment of Switkes et al.~\cite{Switkes} A cyclic modulation
of gate potentials can produce a dc current through an open system.  In contrast
to pumps in the Coulomb-blockade regime,~\cite{LeoK91,Pothier91} pumping through
open systems is governed by quantum effects.  The pumping frequency exceeds the
level spacing violating true adiabaticity,~\cite{Thouless} but a compact
description can still be achieved if the driving fields change weakly on the
scattering time scale.  Brouwer~\cite{Brouwer} expressed the average pumped
charge in terms of the loop traversed by the scattering matrix $S(t)$.  Further
research concentrated on the mesoscopic
fluctuations,~\cite{ZhouSpivakAlt,ShutAleiAlt,PolBrouwer} the role
of discrete symmetries~\cite{ShutAleiAlt,AleiAltKam} and the conditions for
charge quantization,~\cite{ZhouSpivakAlt,ShutAleiAlt,AleiAnd,Levinson} not
necessarily present in open systems.

Statistics of charge transport are of great importance, both for fundamental
research and potential applications.  The counting statistics can be formally
expressed~\cite{LLL} as the determinant of an integral operator involving
$S(t)$.  While compact expressions were obtained for particular pumping
cycles,~\cite{LLL,IvaLev,AndKam} a general case requires further analysis.

Let us begin by sketching our main results.  Our consideration is valid also
in the presence of a voltage bias, since the latter can be gauged away at the
expense of a phase of the scattering matrix.  We will show that the statistics
of current are invariant under a local symmetry group and are determined by 
the path in the corresponding coset space.  Namely, consider the transformation
\begin{equation}
S(t)\to U(t) S(t)
\,,
\qquad
U(t)=\left(\begin{array}{cc}U_L(t)&0\\0&U_R(t)\end{array}\right)
\,,
\label{transform}
\end{equation}
for a scattering matrix $S(t)\in \U(2M)$.
It causes additional rescattering between $M$ left outgoing channels ($U_L$) 
and between $M$ right channels ($U_R$).
We show that $U(t)$ shifts the probability distribution
$P(Q)$ of pumping the charge $Q$ by an integer (we set the elementary charge 
$e=1$),
the relative winding number of the overall phases of $U_L$ and $U_R$:
\begin{equation}
P(Q)\to P(Q-W),\qquad W=\frac{1}{8\pi i}\oint\Tr(dU U^\dagger\sigma_3)
\,.
\label{shift}
\end{equation}
Here $\sigma_3$ is the diagonal matrix $\diag\{1_M,-1_M\}$.
In particular, $P(Q)$ is invariant if the path is trivial, $W=0$.

This result implies that the statistics are determined, up to an integer offset,
by the path $\N(t)=S^\dagger\sigma_3 S$ in the (matrix realization of the) coset
space $\U(2M)/\U(M)\times \U(M)$.  In the single-channel case $\N(t)=\n\bsigma$
reduces to a contour ${\cal C}=\{\n(t)\}$ on the unit sphere in 3D.  We find
that at low temperature the average pumped charge~\cite{Ncycles} is given by the 
area enclosed
by $\cal C$ on the sphere (cf.~Ref.~\onlinecite{Avron}),
\begin{equation}
\left\langle Q \right\rangle = \frac{1}{4\pi}A_{\rm sphere}
\,.
\label{Qav=area}
\end{equation}
This area is defined only modulo $1$ since a surface with the edge $\cal C$ can
cover any of two complementary pieces of the sphere or even cover the sphere
several times.  A way of fixing this integer is discussed below.  When the
contour $\N(t)$ [but not necessarily $S(t)$] is small, the pumped charge is 
quantized.  Indeed, in Refs.~\onlinecite{ShutAleiAlt,Levinson} quantization
was found under these conditions.

\begin{figure}
\centerline{\hbox{\psfig{figure=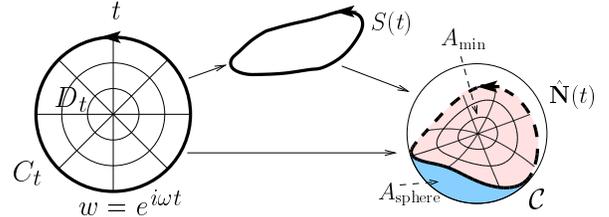,width=0.9\columnwidth}}}
\caption{%
A pumping cycle $S(t)$ defines a contour $\N(t)$ in the 
coset space, which determines the statistics. The area $A_{\rm sphere}$ 
enclosed 
by $\N(t)$ in the coset space and the area $A_{\rm min}$ of the minimal 
surface 
in the space of all matrices define the average pumped charge and the noise.
}
\label{Fig:bundle}
\end{figure}

The current noise was expressed~\cite{AndKam} via a non-local 
integral involving $\N(t)$. We transform it to a local integral over the 
time disk defined as follows: For the analysis of 
driving at frequency $\omega$ we can replace the time axis by a unit circle 
$C_t$: $w=e^{i\omega t}$.
There is a unique harmonic ($\Delta\n=0$) extension of the mapping $\n(w)$ 
from $C_t$ into the disk $D_t$. The noise is given by
the Dirichlet functional,
\begin{equation}
\langle\langle Q^2 \rangle\rangle =
\frac{1}{8\pi} \int_{D_t}(\partial_i\n)^2\, d^2x
\,,
\label{Dirichlet n}
\end{equation}
where $w=x_1+ix_2$.  For a given contour $\cal C$ the details of its traversal
in time depend on the pulse shape of the driving fields.  Unlike the average
charge (\ref{Qav=area}), the noise value is sensitive to the pulse shape, 
i.e.,
to the time parametrization.  Optimal pulse shapes minimizing the noise were
found in several cases.~\cite{LLL,AndKam}  Here we solve the problem of noise
optimization for an arbitrary cycle.  Specifically, we show that the pumping 
cycle is optimal when the mapping $\n(w)$ is conformal, $(\partial_w\n)^2=0$.  
The minimal noise value~\cite{Ncycles} is the area of the minimal surface (soap 
film) spanned 
by $\cal C$,
\begin{equation}
\langle\langle Q^2 \rangle\rangle_{\rm min} = \frac{1}{4\pi} A_{\rm min}
\,.
\label{min noise(n)}
\end{equation}

Similar results hold in the many-channel case.

\paragraph*{Invariance.}

Physically, multiplication of the scattering matrix (\ref{transform}) by
$U(t)$ just redistributes the scattered particles, without affecting the
correlations at the scattering center.  The outgoing states acquire an extra
time-dependent phase which changes the time these particles need to reach the
reservoirs.  As a result, the extra charge $W_i\equiv\oint\Tr(dU_i^\dagger
U_i)/4\pi i$ is transferred to the lead $i=$~L or R, and we get (\ref{shift}).
For periodic $U(t)$ these numbers are integer.  (Note that no net charge
accumulation near the scatterer implies $W_L=-W_R$.)

More formally, the transformation rule (\ref{shift}) for $\langle Q\rangle$ 
follows from the Brouwer formula (\ref{Qaver-1}). For the higher cumulants one 
can use the result~\cite{LLL,AndKam} for the generating function 
$\chi(\lambda) = \sum_Q 
P(Q) e^{i\lambda Q}$,
\begin{equation}
\chi(\lambda)=\det\left[1 + n_F(t',t) 
(S^\dagger_{-\lambda}(t)S_\lambda(t)-1)
\right].
\label{det}
\end{equation}
Here $S_\lambda(t)\equiv e^{-i\lambda\sigma_3/4} S(t) e^{i\lambda\sigma_3/4}$ 
and $n_F(t',t)=i/[2\pi (t'-t+i0)]$ is the Fourier transform of the 
Fermi distribution.
In Ref.~\onlinecite{AndKam}, by separating phases and amplitudes of $S(t)$, 
this result was presented in a form, which immediately implies the 
transformation rule (\ref{shift}). Indeed, the determinant in Eq.~(5) of 
Ref.~\onlinecite{AndKam} is invariant under (\ref{transform}) and the 
quantity $\hat N$ in the prefactor is shifted by $W$.

To express $\chi(\lambda)$ via $\N$ we notice that 
$S^\dagger_{-\lambda}S_\lambda= e^{i\lambda\sigma_3/4} S^\dagger 
e^{-i\lambda\sigma_3/2} S e^{i\lambda\sigma_3/4}= e^{i\lambda\sigma_3/4} 
e^{-i\lambda S^\dagger\sigma_3 
S/2} e^{i\lambda\sigma_3/4}$.
Using the identity $e^{-i\lambda\N/2} = \cos\frac{\lambda}{2}-
i \sin\frac{\lambda}{2}\N$, we get
\begin{equation}
\chi(\lambda)=
\det\left[1 -\case{1}{2}n_F(t'\!,t)
(e^{i\lambda\sigma_3}-1)\sigma_3
[\N(t)-\sigma_3] 
\right].
\label{chi(N)}
\end{equation}
At $T=0$, multiplying by $1+n_F(e^{-i\lambda\sigma_3/2}-1)$, we obtain
\begin{equation}
\chi(\lambda)=\det\left[1+n_F(t'\!,t)(e^{-i\lambda\N(t)/2}-1)\right]
\,.
\label{Eq:chi ort-inv}
\end{equation}
Note that the result (\ref{Eq:chi ort-inv}) is explicitly invariant under 
global rotations $\N(t)\to V^\dagger\N(t) V$ [corresponding to
transformations~\cite{Levitov01} $S(t)\to S(t)V$].

Eqs.~(\ref{det})--(\ref{Eq:chi ort-inv}) involve $\N$, but not $S$, and hence
can define $P(Q)$ only up to an integer offset.  Indeed, the infinite product of
the eigenvalues of these integral operators can be regularized in many
ways.~\cite{IvaLev} Notice that for the operator (\ref{Eq:chi ort-inv}), due to
strong degeneracy, one can choose eigenstates that span a narrow frequency
range, of order $\omega$.  For those far above the Fermi level ($n_F=0$) the
eigenvalues are $1$.  Deep in the Fermi sea ($n_F=1$) the eigenvalues appear in
pairs $e^{\pm i\lambda/2}$ with the product $1$.  Although regularization
procedures can pair them in different ways, $\chi(\lambda)$ can be changed only
by an even power of $e^{\pm i\lambda/2}$, which gives an integer shift of
$Q$.

\paragraph*{Pumped charge and the area.}

The expression for $\langle Q\rangle=\partial_\lambda\chi(\lambda\!=\!0)$ 
contains a singularity whose regularization requires the knowledge of the full 
$S(t)$ and gives an integral over the period~\cite{Brouwer},
\begin{equation}
\langle Q\rangle =
\frac{1}{4\pi i} \oint_C \Tr (\sigma_3 dS S^\dagger)
\,.
\label{Qaver-1}
\end{equation}
The loop $C=\{S(t)\}$ can be contracted to a point, uniquely up 
to continuous deformations. In the process the loop spans a surface $D$. (For 
a two-parametric pump~\cite{Brouwer} there is a natural choice of $D$ 
corresponding to the interior of the contour in the parameter plane.)
Using Stokes' theorem, we rewrite (\ref{Qaver-1}) as a surface integral,
which further reduces to the ``area'' of the corresponding surface $\cal D$ in 
the coset space,
\begin{equation}
\langle Q\rangle =
\int_D
\frac{ \Tr (\sigma_3\, dS \wedge dS^\dagger) }{4\pi i}
=
\int_{\cal D} 
\frac{ \Tr (\N\, d\N\wedge d\N) }{16\pi i}
\,.
\label{Qav(N)}
\end{equation}
Note that the integrand is the curvature of the fiber bundle
$S\to\N=S^\dagger\sigma_3 S$.
In the single-channel case $\N=\n\bsigma$, and we obtain (\ref{Qav=area}):
\begin{equation}
\langle Q\rangle =
\frac{1}{8\pi} \int_{\cal D} \epsilon_{ijk} n_i dn_j \wedge dn_k
\,.
\label{Qav(n)}
\end{equation}
One can try to define the ``integral part'' $Q_{\rm int}$ of $\langle Q\rangle$ 
as follows: Let us parametrize scattering matrices as $S = U S^0[\N]$ with a 
matrix $S^0$, defined for any $\N$, and a matrix $U$ as in 
Eq.~(\ref{transform}), and assign to each cycle $S(t)$ the winding number of the 
corresponding $U(t)$. This attempt fails, since there is no continuous global 
map $S^0[\N]$. In fact, any two loops in $\U(2M)$ can be deformed into each 
other, i.e., any $Q_{\rm int}$ is discontinuous under certain contour 
deformations. However, continuous maps $S^0[\N]$ do exist for contractible 
regions, and one can introduce $Q_{\rm int}$ for contours $S(t)$, for which 
$\N(t)$ does not leave such a region. Examples are regions of matrices $S$ 
without perfectly transmitting (or reflecting) channels. In particular,
the integer $\hat N$ introduced in Ref.~\onlinecite{AndKam} changes abruptly 
when the pumping cycle contains a scattering matrix $S(t)$ with a perfectly
transmitting channel.

At this point we can formulate sufficient conditions for the {\it
quantization} of the pumped charge:  The fractional part of $\langle Q\rangle$
vanishes for small contours $\N(t)$ in the coset space, which remain close to
their initial point over the pumping cycle.  An example:  the minimal and
maximal conductance, $g=0$ and $g=M$, is achieved at the points
$\N=\pm\sigma_3$.  Hence keeping $g$ close to one of these values throughout
the cycle guarantees the quantization.  We evaluate the integral
(\ref{Qav(N)}) for such cycles and estimate the accuracy of quantization as
\begin{equation}
\delta Q\lesssim g \quad\mbox{for } g\approx 0
\,,
\qquad
\delta Q\lesssim M-g \quad\mbox{for } g\approx M
\,.
\end{equation}
For a single channel the fractional part is the area (\ref{Qav=area}) within 
the 
small contour $\n(t)$. 
 
\begin{figure}
\centerline{%
	  \hbox{\psfig{figure=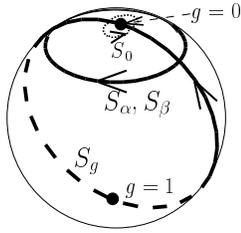,height=0.35\columnwidth}}
}
\caption[]{\label{Fig:sphere}
Several pumping cycles $\n(t)$: $S_\beta$~\cite{LLorig} and 
$S_\alpha$~\cite{AndKam} lead to the same counting statistics, which at 
$g\!=\!1/2$ coincide with that for $S_g$~\cite{AndKam}. The sketched 
contour $S_0$~\cite{ShutAleiAlt,Levinson} describes pumping of a unit charge 
per cycle.}
\end{figure}

For a single channel, $M=1$, the scattering matrix can be parametrized 
by the conductance $g$ and three phases:
\begin{equation}
S(g,\alpha,\beta,\varphi)
=e^{i\varphi/2}
\left(\begin{array}{cc}
\sqrt{1-g}\, e^{i\alpha} & i \sqrt{g}\, e^{i \beta}\\
i \sqrt{g}\, e^{-i \beta} & \sqrt{1-g}\, e^{-i\alpha}
\end{array}\right)
\,.
\label{4param}
\end{equation}
The components of the unit vector $\n$ then are
\begin{equation}
n_z = 1-2g \,;\quad
n_x+i n_y = -2i \sqrt{g(1-g)}\, e^{i(\alpha-\beta)}\,.
\label{n-vector}
\end{equation}
Using these expressions we can explain the charge quantization found, for 
instance, in Refs.~\onlinecite{ShutAleiAlt,Levinson}: For the pumping 
cycles studied the system 
encircled the resonance point $g=1$ in the parameter plane at a sufficient 
distance from it so that $g\approx0$ throughout the cycle. The corresponding 
loop $\n(t)$ encircled the north pole ($g=0$; Fig.~\ref{Fig:sphere}). Since 
the 
interior of the loop in the parameter plane contained the resonance point, the 
surface $\cal D$ in (\ref{Qav(n)}) covered the {\em lower} part of the sphere, 
i.e., almost the whole sphere, $\langle Q\rangle\approx1$.

\paragraph*{Noise optimization.}

For the noise, given by the $\lambda^2$ term in the Taylor series of
$\ln\chi(\lambda)$, we obtain a double integral~\cite{LLL,AndKam} over the unit
circles $w=e^{i\omega t}$, $w'=e^{i\omega t'}$:
\begin{equation}
\langle\langle Q^2 \rangle\rangle =
\frac{1}{16\pi^2}\oint\!\!\oint
\frac{dw\, dw'}{(w-w')^2}\,
\Tr\,[1 - \,\N(t)\N(t')\,]
\,.
\label{noise}
\end{equation}
Let $\N(t)=\sum_{k\ge 0} \N_k \exp(i k\omega t) + \mbox{h.c.}$ Then the (unique) 
harmonic extension of the mapping $w\to\N(w)$ from the circle
into the disk $|w|<1$ is given by
\begin{equation}
\N(w) = \N^+(w) + \N^-(w) = \sum\nolimits_{k\ge0} \N_k w^k +\mbox{h.c.}
\label{N=N+ + N-}
\end{equation}

Expanding the integrand (\ref{noise}) near $t'=t$ one sees that we are 
justified in replacing the integration over $w$ by
\begin{equation}
\oint\nolimits_{|w|=1} \to \frac{1}{2}\oint\nolimits_{|w|=1-\varepsilon}
+ \frac{1}{2} \oint\nolimits_{|w|=1+\varepsilon}
\,,\qquad \varepsilon\to0
\,.
\label{regul}
\end{equation}
Eqs.~(\ref{N=N+ + N-}), (\ref{regul}) and the invariance under time reversal 
$w,w'\to\wbar,\wbar'$ allow us to use complex analysis to do the integration 
over $w'$ in (\ref{noise}):
$$
\langle\langle Q^2 \rangle\rangle =
\Tr
\oint \frac{\N(w)}{16\pi i}
\left[
dw\, \partial_w \N^+(w)
-
d\wbar\, \partial_\wbar \N^-(w)
\right]
\,.
$$
Finally, Stokes' theorem gives the integral over the interior 
$D_t=\{w=x_1+ix_2: |w|\le1\}$ of the time circle,
\begin{equation}
\langle\langle Q^2 \rangle\rangle =
\frac{1}{16\pi}
\int_{D_t} d^2x \Tr[(\partial_i\N)^2]
\,.
\label{Dirichlet N}
\end{equation}
which reduces to (\ref{Dirichlet n}) in the single-channel case.
Note that the functional (\ref{Dirichlet N}) is well-defined for any surface 
$\N(w)$. The harmonic surface (\ref{N=N+ + N-}) provides the minimal value to 
(\ref{Dirichlet N}) among the maps $\N(w)$ with the fixed value at the 
boundary, 
$\N(t)$.

Now we turn to optimization of pumping:  The cyclic evolution of $S(t)$ is
achieved by periodic changes in external parameters that control the
scattering or the bias.  For a fixed trajectory in the parameter space the
rate of motion can be varied.  These changes do not affect the average pumped
charge (\ref{Qaver-1}), (\ref{Qav(n)}) but do influence the noise
(\ref{Dirichlet N}).  Notice that~\cite{DNF} $\frac{1}{2} \Tr[(\partial_1\N)^2
+ (\partial_2\N)^2] \ge \sqrt{\Tr[(\partial_1\N)^2] \cdot
\Tr[(\partial_2\N)^2]} \ge \mbox{area spanned by }\partial_1\N\mbox{ and
}\partial_2\N$ in the matrix space.  Hence the noise value always exceeds the
area [defined by the scalar product $\langle A,B\rangle \equiv \Tr(A^\dagger
B)/2$]:
\begin{equation}
\langle\langle Q^2 \rangle\rangle \ge
\frac{1}{4\pi} A[\N(w)]
\,.
\label{ineq}
\end{equation}
The equality is achieved only if $\partial_i\N$ are orthogonal,
$\Tr(\partial_1\N\partial_2\N)=0$, and have the same length,
$\Tr[(\partial_1\N)^2]=\Tr[(\partial_2\N)^2]$.  These two conditions together
can be written as $\Tr[(\partial_w\N)^2]=0$ and characterize {\it conformal}
mappings $\N(w)$.

Since any regular surface can be parametrized conformally, the minimal values
of both sides of Eq.~(\ref{ineq}) coincide.  The minimal noise value is thus
given by the {\it minimal area} (\ref{min noise(n)}) of a surface spanned by
the loop $\N(t)$.  The optimal pumping with this minimal noise value for a
given loop is achieved when the corresponding harmonic mapping $\N(w)$ is
conformal.  Notice that the $\SL_2(\mathbb{R})$ time reparametrization
symmetry~\cite{IvaLev}, $w\to(w+a)/(1+\bar a w)$, preserves the classes of
harmonic and conformal maps.  For $N$ cycles $\N_N(w)\!\!\equiv\!\!\N_1(w^N)$
and $\N_1(\prod_{i=1}^N (w+a_i)/(1+\bar a_i w))$ give the same statistics.

\paragraph*{Applications.}
 
Our findings give a new perspective on the analysis of pumping cycles 
discussed in the literature.
Consider the cycles
\begin{eqnarray}
S_\beta(t) &=& e^{-i\phi(t)\sigma_3/2} S(0) e^{i\phi(t)\sigma_3/2}
\,,
\quad_{\displaystyle\Delta\phi=2\pi N,}
\label{Eq:beta(t)}
\\[-1mm]
S_\alpha(t) &=& \phantom{^-} e^{i\phi(t)\sigma_3/2} S(0) 
e^{i\phi(t)\sigma_3/2}
\,.
\label{Eq:alpha(t)}
\end{eqnarray}
The first of them, studied extensively by Levitov et al.,~\cite{LLorig,LLL}
describes conductors under the voltage bias $-\hbar\dot\phi(t)/e$, as one can
see by applying a gauge transformation.  The cycle $S_\alpha$ was discussed in
Ref.~\onlinecite{AndKam}.  These cycles differ only by a transformation
(\ref{transform}) with $U(t)=e^{i\phi(t)\sigma_3}$, hence the statistics
coincide up to a shift by the winding number:  $P_\beta(Q-N) = P_\alpha(Q)$.
Indeed, the same pulse shape $\phi(t)=\omega t$ [and others, generated by
$\SL_2(\mathbb{R})$] was found optimal for both cycles.  The statistics for this
optimal cycle $S_\alpha(t)$ are related to the well-known binomial distribution
for a conductor under a constant {\em positive}~\cite{sign-note} bias by $P^{\rm
opt}_\beta(N-Q)=P^{\rm opt}_\alpha(Q)$, in agreement with
Ref.~\onlinecite{AndKam}.  For a single channel the vector $\n(t)$ follows a
line of constant latitude for both $S_\alpha$ and $S_\beta$:
$\beta(t)-\beta(0)=\phi(t)$ or $\alpha(0)-\alpha(t)=\phi(t)$.  The pumped charge
$\langle Q\rangle$ is given by the area above this line, $g$, for
(\ref{Eq:beta(t)}), and below this line, $1-g$, for (\ref{Eq:alpha(t)}).  The
minimal noise value (\ref{min noise(n)}) is the area $g(1-g)$ of the sphere's
cross-section.  The optimal pumping corresponds to the trivial homothety of the
time disk onto this cross-section, this map being obviously harmonic and
conformal.

The rotational invariance of $\chi(\lambda)$ implies that the counting 
statistics for any circular path $\n(t)$ is the same as for biased 
conductors.~\cite{LLL} In particular, the optimal pumping 
uniformly traverses the circle and gives rise to a binomial 
distribution.
As an example consider the cycle
\begin{equation}
S_g(t)=
\left(\begin{array}{cc}
\cos\eta(t) & \sin\eta(t)\\
\sin\eta(t) & -\cos\eta(t)
\end{array}\right)
\label{Eq:eta(t)}
\,,
\qquad
\Delta\eta=2\pi
\,,
\end{equation}
during which the conductance $g=\sin^2 \eta(t)$ oscillates. We find that
$\langle Q\rangle=0$ and $\n(t)$ traverses twice the meridian in 
Fig.~\ref{Fig:sphere}. Thus $P(Q)$ coincides with the distribution for 
the equator, $S_\beta$ at $g=1/2$, and the pulse $\eta(t)=\omega t$ 
from Ref.~\onlinecite{AndKam} is optimal. For this pulse $P(Q)$
is the shifted binomial distribution for $N=2$ cycles,  $P^{\rm 
opt}_g(Q)=P^{\rm opt}_\beta(Q+1)\big|_{N=2}^{g=1/2}=\frac{1}{4}{2\choose 
Q+1}$, in agreement with Ref.~\onlinecite{AndKam}.

Our geometric approach allows us to obtain relations between current and noise
for broad classes of pumping cycles.  For small loops $\n(t)$ the minimal
surface lies within the sphere and the analysis simplifies.  If the loop $S(t)$
is also small, the system is in the weak pumping regime with $\langle Q\rangle,
\langle\langle Q^2\rangle\rangle\ll 1$.  In this case we have for a general
(possibly self-intersecting) loop $\langle Q\rangle=(A_+ - A_-)/4\pi \le
\langle\langle Q^2\rangle\rangle_{\rm min} =(A_+ + A_-)/4\pi$, where $A_\pm$ are
the contributions to the enclosed area with positive (resp.  negative)
orientations.  The weak-pumping regime was studied very recently by
Levitov~\cite{Levitov01} who found that the transport is described by two
uncorrelated Poisson processes (transporting charge to the right and to the
left), reducing in some cases to a single process.  Our inequality
$\langle\langle Q^2\rangle\rangle \ge \langle Q\rangle$ is in agreement with
Levitov's findings, with $A_\pm/4\pi$ being the rates of the two Poisson
processes for an optimal cycle.  The equality, the criterion for the reduction
to a single Poisson process in the weak-pumping regime, is thus reached only for
optimally traversed contours enclosing the area of a constant orientation (in
particular, for non-self-intersecting loops).  The simplest example of such a
cycle considered in Ref.~\onlinecite{Levitov01} corresponds in our terms to
${\bf n}(t)$ traversing uniformly a small circle.
Generally, for weak harmonic driving $\n(t)$ encircles an ellipse. The optimal 
pulse shape, given by the conformal map of $D_t$ onto this ellipse, involves 
elliptic integrals.~\cite{Fuks} Further, for a general small polygon the optimal 
pumping is given by the Schwarz-Christoffel formula, describing a map of $D_t$ 
onto its interior (also reducing to elliptic 
integrals for a rectangle).~\cite{Fuks}

The results concerning the weak-pumping regime can be generalized to the
many-channel case.  Using local complex coordinates in the coset space, we
found that the ratio $\langle\langle Q^2\rangle\rangle/|\langle Q\rangle| \ge
1$, reaching the minimal value unity (corresponding to a single Poisson
process~\cite{Levitov01}) only for optimal cycles with a complex analytic (or
antianalytic) minimal surface $\N(w)$.

For the strong-pumping regime our description also gives new results.  For
instance, in the interesting case of a single channel and a contour $\n(t)$
without self-intersection, we find that $\langle\langle Q^2\rangle\rangle_{\rm
min}\le$ distance from $\langle Q\rangle$ to closest integer.

In conclusion, we have shown that the counting statistics of a sample subject
to a periodic pumping (and possibly to a voltage bias) are determined by a
path in the coset space and given by Eq.~(\ref{Eq:chi ort-inv}).  The average
pumped charge and its minimal variance for an arbitrary pumping cycle are
given by the area (\ref{Qav(N)}) encircled by this path in the coset space and
the area (\ref{min noise(n)}) of the minimal surface spanned by this path,
respectively.  The optimal pumping strategy can be found as a harmonic
conformal map of the time disk onto this minimal surface.  Our results
represent a unifying framework for analysis of transport statistics in various
realizations of pumping.

We acknowledge support from the SFB195 of the DFG.


\begin{thebibliography}{99}
\bibitem{Ncycles}
The results for $\langle Q\rangle$, $\langle\langle Q^2\rangle\rangle$, $\ln 
\chi$ are given per cycle, i.e., should be multiplied by the number $N\gg1$ of 
cycles.
\end{thebibliography}
\end{document}